\documentclass[
    ,final
  ]
  {aipproc}

\layoutstyle{6x9}

\newcommand\doingARLO[2][]{%
  \ifx\mmref\undefined #1\else #2\fi
}

\begin{document}

\title 
      [Are Short GRBs Really Hard?]
      {Are Short GRBs Really Hard?}


\author{T. Sakamoto}{
  address={NASA Goddard Space Flight Center},
  altaddress={National Research Council},
  email={takanori@milkyway.gsfc.nasa.gov},
  thanks={}
}

\iftrue
\author{L. Barbier}{
  address={NASA Goddard Space Flight Center},
}

\author{S. Barthelmy}{
  address={NASA Goddard Space Flight Center},
}

\author{J. Cummings}{
  address={NASA Goddard Space Flight Center},
  altaddress={National Research Council},
}

\author{E. Fenimore}{
  address={Los Alamos National Laboratory},
}

\author{N. Gehrels}{
  address={NASA Goddard Space Flight Center},
}

\author{D. Hullinger}{
  address={NASA Goddard Space Flight Center},
  altaddress={University of Maryland},
}
\author{H. Krimm}{
  address={NASA Goddard Space Flight Center},
  altaddress={Universities Space Research Association},
}

\author{C. Markwardt}{
  address={NASA Goddard Space Flight Center},
  altaddress={Universities Space Research Association},
}

\author{D. Palmer}{
  address={Los Alamos National Laboratory},
}

\author{A. Parsons}{
  address={NASA Goddard Space Flight Center},
}

\author{G. Sato}{
  address={Institute of Space and Astronautical Science},
}

\author{J. Tueller}{
  address={NASA Goddard Space Flight Center},
}

\author{R. Aptekar}{
  address={Ioffe Physico-Technical Institute},
}

\author{T. Cline}{
  address={NASA Goddard Space Flight Center},
}

\author{S. Golenetskii}{
  address={Ioffe Physico-Technical Institute},
}

\author{E. Mazets}{
  address={Ioffe Physico-Technical Institute},
}

\author{V. Pal'shin}{
  address={Ioffe Physico-Technical Institute},
}

\author{G. Ricker}{
  address={Massachusetts Institute of Technology},
}

\author{D. Lamb}{
  address={University of Chicago}
}

\author{J.-L. Atteia}{
  address={Observatoire Midi-Pyren$'{e}$es}
}

\author{N. Kawai}{
  address={Tokyo Institute of Technology}
}

\author{Swift-BAT}{
  address={}
}

\author{Konus-Wind}{
  address={}
}

\author{HETE-2 team}{
  address={}
}

\fi

\copyrightyear  {2001}

\begin{abstract}
Thanks to the rapid position notice and response by HETE-2 and Swift, 
the X-ray afterglow emissions have been found for four recent short 
gamma-ray bursts (GRBs; GRB 050509b, GRB 050709, GRB 050724, and GRB 050813). 
The positions of three out of four short GRBs are coincident with galaxies 
with no current or recent star formation. This discovery tightens the case 
for a different origin for short and long GRBs.
On the other hand, from the prompt emission point of view, a short GRB shows 
a harder spectrum comparing to that of the long duration GRBs according to 
the BATSE observations. We investigate the prompt emission properties of 
four short GRBs observed by Swift/BAT. We found that the hardness of all 
four BAT short GRBs is in between the BATSE range for short and long GRBs.
We will discuss the spectral properties of short GRBs including the short 
GRB sample of Konus-Wind and HETE-2 to understand the hard nature of the 
BATSE short GRBs. 
\end{abstract}

\date{\today}

\maketitle

\section{Introduction}

In the year of 2005, there is a major progress in understanding 
of the nature of short GRBs.  $Swift$ X-Ray Telescope (XRT) found 
the X-ray afterglows from the short GRBs; GRB 050509B \citep{gehrels2005}, 
GRB 050724 \citep{barthelmy2005} and GRB 050813 \citep{retter2005,
morris2005} detected by $Swift$ Burst Alert Telescope (BAT).  This 
discovery allows us to pinpoint the location of the short GRBs within 
10$^{\prime\prime}$ for the first time.  And also the short GRB, 
GRB 050709, observed by $HETE$-2 allows us to determine the position less 
then 1$^{\prime\prime}$ thanks to the follow-up observations by HST and 
$Chandra$ \citep{fox2005}.  
Because of these accurate position measurements, we start to understand 
that the environment and/or the progenitor of short GRBs might be 
different from the long GRBs (e.g. \citep{prochaska2005}).  

From the prompt emission point of view, one of the most popular 
characteristic of short GRBs is the spectral hardness of these bursts.  
The left of figure \ref{fig:dur_hardness1} shows the T$_{90}$ duration 
versus the fluence ratio between 100--300 (BATSE channel 3) and 50--100 
(BATSE channel 2) keV band (hereafter HR32) for the BATSE GRBs 
\citep{paciesas1999}.  As see in this figure, the 
short GRBs (T$_{90}$ $<$ 2 s) have a significantly larger hardness ratio 
comparing to that of the long GRBs.  About 48\% and 26\% of the BATSE short 
GRBs having HR32 grater than 6 and 8 respectively.  

In this paper, we will focus on the prompt emission spectral properties of 
short GRBs observed by $Swift$/BAT, $HETE$-2, and $Konus$-$Wind$ to 
investigate the hardness of the short GRBs.  

\section{T$_{90}$ - Hardness ratio}

\begin{figure}[t]
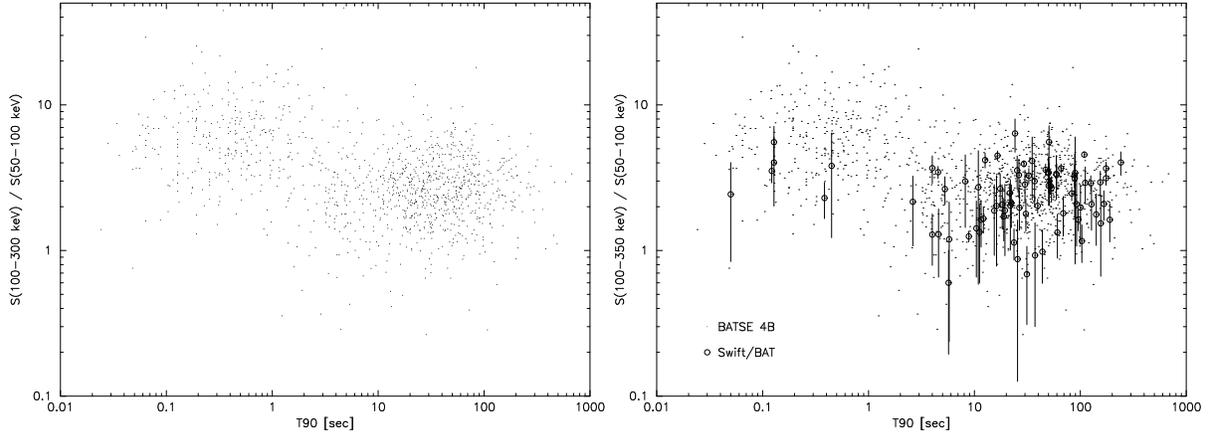

  {\includegraphics[scale=.33,angle=-90]{fig1a.eps}
  \includegraphics[scale=.33,angle=-90]{fig1b.eps}}
  \caption{T$_{90}$ - Hardness plot (left: BATSE and right: BAT).  
The error bar in the BAT sample is 90\% confidence level.}
 \label{fig:dur_hardness1}
\end{figure}

\begin{figure}[h]
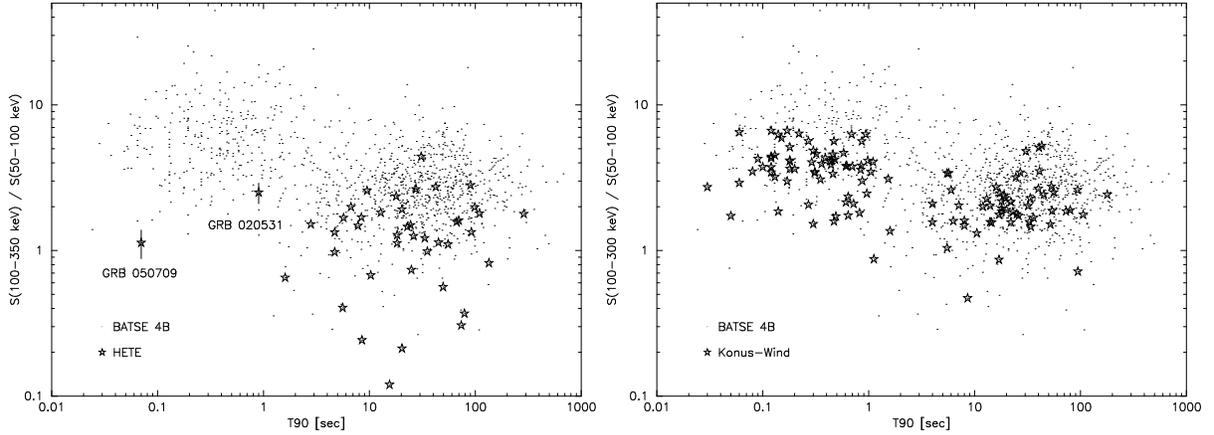

  {\includegraphics[scale=.33,angle=-90]{fig2a.eps}
  \includegraphics[scale=.33,angle=-90]{fig2b.eps}}
  \caption{T$_{90}$ - Hardness plot (left: HETE and right: Konus-Wind).  
Note that the reason for many long GRBs showing the smaller number of
 HR32 in the 
$HETE$-2 GRB sample is that $HETE$-2 is observing large number of 
``soft'' GRBs, so called XRFs \citep{sakamoto2005}.}
 \label{fig:dur_hardness2}
\end{figure}

Figure \ref{fig:dur_hardness1} and \ref{fig:dur_hardness2} show 
T$_{90}$ vs. HR32 for the GRB sample of BATSE, BAT, $HETE$-2, and 
$Konus$-$Wind$.  Although the HR32 for the long GRBs are consistent 
with the BATSE GRB sample, HR32 for the short GRBs observed by BAT, 
$HETE$-2 and $Konus$-$Wind$ is not as hard as the BATSE short GRBs.  
There are about a quarter of the BATSE short GRBs having the HR32 
grater than 8.  However, none of the short GRBs observed by 
BAT, $HETE$-2 and $Konus$-$Wind$ having the similar amount of HR32.  

\begin{figure}[h]
  {\includegraphics[scale=.33,angle=0]{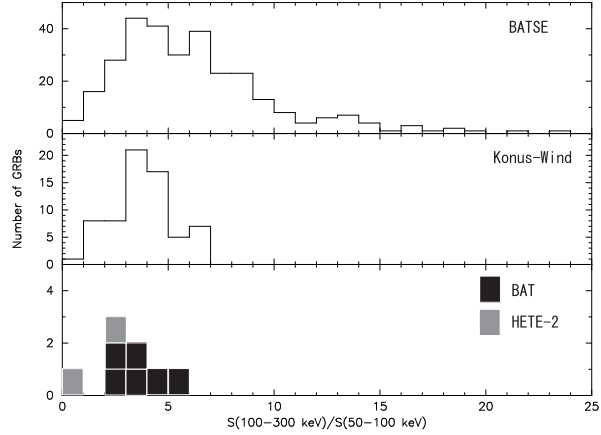}}
  \caption{The histogram of HR32 for the short GRBs observed by 
BATSE, Konus-Wind, HETE-2, and Swift/BAT.}
  \label{fig:hist_hardness_summary}
\end{figure}

HR32 of the short GRBs of all four GRB instruments is summarized in 
figure \ref{fig:hist_hardness_summary}.  Although the number of the sample is 
limited for BAT and $HETE$-2, the distribution of HR32 for short GRBs 
observed by $Konus$-$Wind$, $HETE$-2, and BAT are all consistent.  
However, the BATSE HR32 distribution is spread over to much wider range, 
especially for the larger value of HR32.    

\begin{figure}[h]
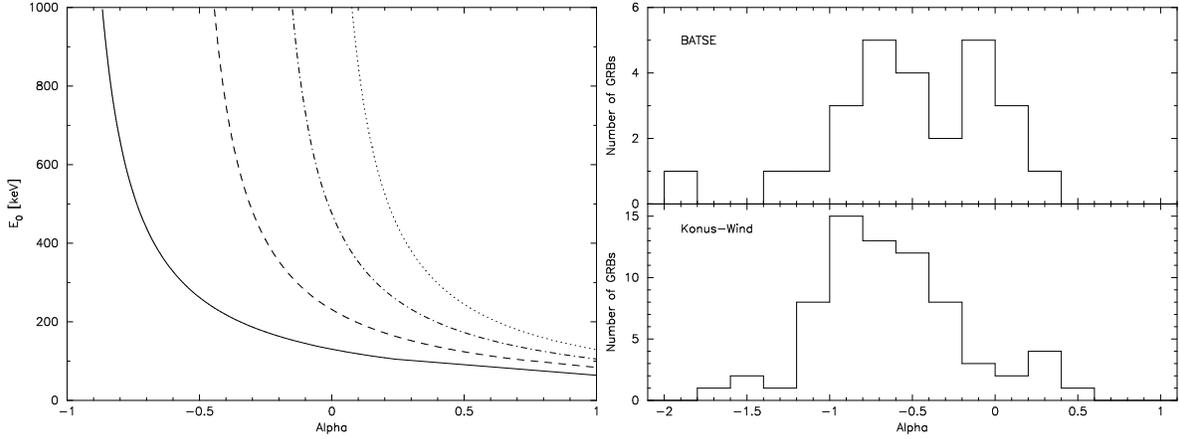

  {\includegraphics[scale=.33,angle=-90]{fig4a.eps}
   \includegraphics[scale=.33,angle=-90]{fig4b.eps}}
   \caption{Left: The calculated $\alpha$ and $E_{0}$ as a function 
of HR32 (solid, dashed, dash-dotted, and dotted lines are HR32 of 4, 6, 8, 
and 10 respectively.  Right: The photon index $\alpha$ distribution 
of the BATSE 
\citep{ghirlanda2004} and the Konus-Wind \citep{mazets2005} short GRB 
sample.}
   \label{fig:alpha_e0}
\end{figure}

\section{Why BATSE short GRBs look harder?}

To investigate the origin of the hardness seen in the BATSE short GRBs, 
we calculate HR32 with a power-law times exponential cutoff 
model\footnote{dN/dE $\sim$ $E^{\alpha}$ $\exp(-E/E_{0}$)} as a 
function of the power-law index, $\alpha$, and the cutoff energy, 
$E_{0}$.  The result is shown in the left panel of figure 
\ref{fig:alpha_e0}.  
To archive HR32 $>$ 6 in a cutoff power-law model, $\alpha$ has to be grater 
than 0.  If $E_{0}$ is less than 500 keV, $\alpha$ should 
increase rapidly when $E_{0}$ goes smaller.  Thus, we may 
conclude that a large value of HR32 seen in the BATSE short GRBs is as a 
result of $\alpha$ $>$ 0, but not of a large $E_{0}$ energy.  

To check our hypothesis, we made the comparison of $\alpha$ distribution 
between BATSE \citep{ghirlanda2004} and Konus-Wind \citep{mazets2005} sample.  
The right panel of figure \ref{fig:alpha_e0} shows the comparison.  
The $\alpha$ distribution of the Konus-Wind short GRB sample has a 
tight distribution which is centroid around $-0.8$.  On the other hand, 
as mentioned by \citet{ghirlanda2004}, a large fraction of the BATSE
short GRBs has $\alpha$ $>$ 0.  This result tightens our conclusion that  
the hardness of the BATSE short GRBs are coming from the extremely flat 
photon index $\alpha$.  

\begin{figure}[t]
  {\includegraphics[scale=.33,angle=-90]{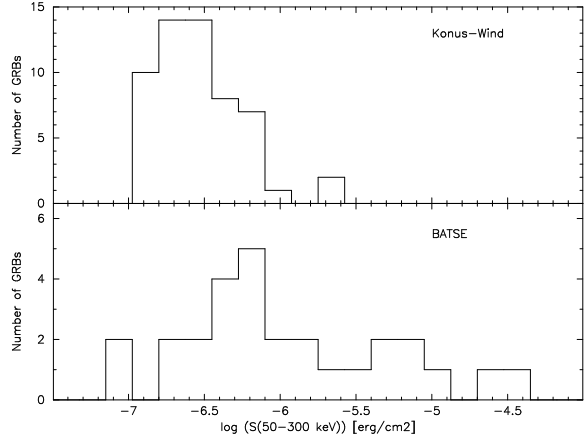}}
   \caption{The 50-300 keV fluence distribution of the Konus-Wind and 
BATSE short GRB sample.  Note that in the Konus-Wind short GRB sample of 
this figure, we only include the short GRBs with the time interval of 
the spectrum less than 256 ms.  Thus, we are not including the large 
fluence bursts due to the difficulty in calculating the fluence in the 
50-300 keV band.}
   \label{fig:fluence_distribution}
\end{figure}

As shown in figure \ref{fig:fluence_distribution}, the BATSE short GRBs 
studied by \citet{ghirlanda2004} are the bright short GRBs.  Since the 
Konus-Wind short GRBs studied 
by \citet{mazets2005} are well covering the fluence range of the BATSE 
short GRB sample, it is difficult to understand the systematic difference 
in $\alpha$ distribution between Konus-Wind and BATSE by the selection 
effect of the sample.  However, we need the complete spectral
information of the BATSE short GRBs to confirm our conclusion.  

We study the hardness of the prompt emission of short GRBs observed by 
four different GRB instruments.  We found that the hardness ratios of 
the short GRBs observed by BATSE have a systematically larger value than 
the short GRBs observed by other instruments.  We also confirmed that 
the hardness of the BATSE short GRBs is as a result of the extremely 
flat power-law index $\alpha$ ($\alpha$ $>$0) which is not a dominant 
population in the $Konus$-$Wind$ short GRB sample.

\end{document}